\documentclass[pre,twocolumn,amsmath,amssymb,nofootinbib,floatfix,superscriptaddress]{revtex4}

\usepackage{enumerate}
\usepackage{graphicx,bm}
\usepackage[normalem]{ulem} 

\makeatletter
\def\graphicscale{\twocolumn@sw{0.3}{0.4}}
\def\graphicthreescale{\twocolumn@sw{0.3}{0.4}}

\begin{document}

\title{Critical behaviors of lattice U(1) gauge models and \\
three-dimensional Abelian-Higgs gauge field theory}

\author{Claudio Bonati} 
\affiliation{Dipartimento di Fisica dell'Universit\`a di Pisa
        and INFN Largo Pontecorvo 3, I-56127 Pisa, Italy}

\author{Andrea Pelissetto}
\affiliation{Dipartimento di Fisica dell'Universit\`a di Roma Sapienza
        and INFN Sezione di Roma I, I-00185 Roma, Italy}

\author{Ettore Vicari} 
\affiliation{Dipartimento di Fisica dell'Universit\`a di Pisa
        and INFN Largo Pontecorvo 3, I-56127 Pisa, Italy}

\date{\today}

\begin{abstract}
We investigate under which conditions the three-dimensional (3D)
multicomponent Abelian-Higgs (AH) field theory (scalar
electrodynamics) is the continuum limit of statistical
lattice gauge models, i.e., when it characterizes the universal 
behavior at critical transitions occurring in these models.
We perform Monte Carlo simulations of the lattice AH model with
compact gauge fields and $N$-component scalar fields  with charge $q\ge 2$
for $N=15$ and 25. Finite-size scaling analyses
of the Monte Carlo data show that the transitions along the
line separating the confined
and deconfined phases are continuous and that they
belong to the same universality class for
any $q\ge 2$. Moreover, they are in the same universality class as the 
transitions in the lattice AH model with noncompact gauge fields along the
Coulomb-to-Higgs transition line. We finally argue that these critical
behaviors are described by the stable charged fixed point of the
renormalization-group flow of the 3D AH field theory.
\end{abstract}

\maketitle


\section{Introduction}
\label{intro}

Three-dimensional (3D) Abelian U(1) gauge models with multicomponent
scalar fields and SU($N$) global symmetry ($N\ge 2$)---the 
Abelian-Higgs (AH) models---emerge in many physical situations. 
They provide effective
theories for superconductors, superfluids, and quantum SU($N$)
antiferromagnets~\cite{RS-90, TIM-05, TIM-06, Kaul-12, KS-12, BMK-13,
  NCSOS-15, WNMXS-17}. In particular, they are expected to describe the
transition between the N\'eel and the valence-bond-solid state in
two-dimensional antiferromagnetic SU(2) quantum
systems~\cite{Sandvik-07, MK-08, JNCW-08, Sandvik-10, HSOMLWTK-13,
  CHDKPS-13, PDA-13, SGS-16}, which represents the paradigmatic model
for the so-called deconfined quantum criticality~\cite{SBSVF-04}. In
this context several studies have focused on systems with two scalar
components ~\cite{MS-90,MV-04,SBSVF-04,KPST-06, Sandvik-07, MK-08,
  JNCW-08, MV-08, KMPST-08-a, KMPST-08, CAP-08, LSK-09, CGTAB-09,
  CA-10, BDA-10, Sandvik-10, HBBS-13, Bartosch-13, HSOMLWTK-13,
  CHDKPS-13, PDA-13, BS-13, NCSOS-15, NSCOS-15, SP-15, SGS-16,
  PV-19-AH3d, PV-20-mfcp, SN-19, SZ-20}.

Classical and quantum Abelian models have 
been extensively studied with the purpose of identifying their
phases and the nature of their phase transitions. It has been
realized that a crucial role is played by topological aspects,
like Berry phases, monopoles, or the
compact/noncompact nature of the U(1) gauge fields, together with the
charge of the scalar fields. Indeed, 
the phase diagram and the nature of the transitions is different in
lattice AH models with compact and noncompact
gauge fields~\cite{PV-19-AH3d,BPV-21-ncAH}, in AH compact models 
with charge-one and higher-charge scalar fields 
~\cite{PV-19-AH3d,BPV-20-hcAH}, and in models
with or without topological defects such as
monopoles~\cite{MS-90,MV-04,PV-19-CP,PV-20-mfcp}.

Multicomponent lattice AH models with U(1) gauge invariance and
SU($N$) global symmetry are the lattice counterparts of 
the multicomponent scalar electrodynamics or AH field theory, in which an
$N$-component complex scalar field ${\bm \Phi}({\bm x})$ is minimally
coupled to the electromagnetic field $A_\mu({\bm x})$. The
corresponding continuum Lagrangian reads
\begin{equation}
{\cal L} = 
|D_\mu{\bm\Phi}|^2
+ r\, {\bm \Phi}^*{\bm \Phi} + 
\frac{1}{6} u \,({\bm \Phi}^*{\bm \Phi})^2 + 
\frac{1}{4 g^2} \,F_{\mu\nu}^2  \,,
\label{AHFT}
\end{equation}
where $F_{\mu\nu}\equiv \partial_\mu A_\nu - \partial_\nu A_\mu$, and
$D_\mu \equiv \partial_\mu + i A_\mu$.  Its renormalization-group (RG)
flow was investigated perturbatively, using the
$\varepsilon\equiv 4-d$ expansion
~\cite{HLM-74,FH-96,IZMHS-19,WK-74,Fisher-75}, in the functional
RG~\cite{FH-17} and in the large-$N$
approach~\cite{HLM-74,DHMNP-81,YKK-96,MZ-03,KS-08}.  These studies
showed that a stable charged fixed point (CFP) with a nonzero gauge
coupling exists only when the number $N$ of components is larger
than $N_D^*$,
where $N_D^*$ depends on the space dimension $D$.
Close to four dimensions, a stable CFP exists only in systems with a 
very large number of components, since
$N_4^* = 90 + 24\sqrt{15} \approx 183$. However, $N_D^*$
drastically decreases 
in three dimensions, $N_3^* \ll N_4^*$.  The 3D value
$N_3^*$ has been estimated by constrained resummations of the
four-loop $\varepsilon$ expansion using two-dimensional
results~\cite{IZMHS-19}, obtaining $N_3^* = 12(4)$, and from 
the analysis of Monte Carlo results for the 
noncompact lattice AH model~\cite{BPV-21-ncAH}, obtaining
$N_3^*=7(2)$.

On general grounds, one would expect
the stable CFP of the 3D RG flow of the AH field theory
to be associated with the universality class of critical transitions
in 3D systems with local U(1) and global SU($N$) symmetry.
However, the behavior of lattice AH models is not so simple.
Indeed, they present different phases and
transitions belonging to different
universality classes, depending on features that are not present 
in the continuum field theory. 
At present, we do not yet satisfactorily understand
under which conditions statistical models  have transitions
controlled by the field-theory CFP. In
particular, it is not clear which are the key features of 
those lattice U(1) gauge models that have critical transitions 
described by the field-theory CFP of the RG flow of the AH field theory.

In general, critical transitions in lattice gauge theories can be 
classified in two different groups:
\begin{enumerate}[i)]
\item Transitions in which only matter correlations are critical;
  at the transition gauge variables do not display long-range correlations.
\item Transitions in which matter and gauge-field correlations are both 
  critical.
\end{enumerate}
In case (i), although gauge variables are not critical,
the gauge symmetry is crucial for identifying the scalar
critical degrees of freedom.  Indeed, gauge symmetry prevents non-gauge
invariant correlators from acquiring nonvanishing vacuum expectation
values and developing long-range order: the gauge symmetry hinders
some scalar degrees of freedom---those that are not gauge
invariant---from becoming critical.  In this case the critical
behavior or continuum limit is driven by the condensation of
gauge-invariant scalar operators, which play the role of fundamental
fields in the Landau-Ginzburg-Wilson (LGW) theory that provides an
effective description of the critical regime, without including the
gauge fields.  The lattice CP$^{N-1}$ model is an example of a U(1)
gauge model that shows this type of behavior~\cite{PV-19-CP,
  PV-20-largeNCP}. Two-dimensional
U(1) gauge models with multicomponent scalar
matter~\cite{BPV-20-2dAH} and several lattice nonabelian gauge
Higgs models in two and three dimensions~\cite{BPV-19-sqcd,
  BFPV-21-sunfu, BFPV-21-sunadj, BPV-20-son, BPV-20-sun2d,
  BFPV-20-on2d, BFPV-21-sunadj2d} also belong to class (i).

In case (ii), in which  both scalar 
and gauge correlations are critical at the transition,
an appropriate effective field-theory description of the critical
behavior requires explicit gauge fields.
Therefore, one would expect that the field-theory CFP of the 
RG flow of the AH field theory is the one that controls
the universal features of the 
critical transitions of type (ii) in AH lattice models.
Critical behaviors consistent with the 
universality classes of the AH field theory have been observed  in
the lattice AH model with noncompact gauge
fields~\cite{BPV-21-ncAH} (along the transition line that separates the 
Coulomb and the Higgs phase), and
in the lattice AH model with compact gauge fields and $q=2$
scalar charge~\cite{BPV-20-hcAH} (along the  transition line between 
the confined and the deconfined phase). We also mention that continuous
transitions of type (ii) have been observed in a different lattice 
U(1) gauge model, in the CP$^{N-1}$ model without
monopoles~\cite{PV-20-mfcp}.  However, they do not belong to the same
universality class as those observed in the noncompact lattice AH
model~\cite{BPV-21-ncAH}.
  
In this paper we return to this issue, strengthening  previous results.
We provide compelling numerical evidence that,
for a sufficiently large number of components $N$, $N\gtrsim 10$, say,
the continuous transitions between the confined and deconfined phase
of the lattice AH model with compact gauge fields and scalar charge $q\ge
2$ belong to the same universality class for any $q\ge 2$. Moreover, 
the critical behavior is the same as in the noncompact AH model, which 
is formally obtained in the limit $q\to\infty$. A detailed 
finite-size scaling (FSS) analysis of the Monte Carlo (MC) results  
allows us to obtain precise estimates of the critical exponents. 
They turn out to be in excellent agreement with the 
field-theory predictions, obtained in the large-$N$ expansion
~\cite{HLM-74,YKK-96,MZ-03}. Therefore, we conclude that the CFP of the 
AH field theory is associated with a line of 
critical transitions that is present
in the lattice AH model with compact gauge fields and 
any scalar charge $q\ge 2$ and in the model with noncompact gauge fields.
In all cases, the field-theory critical behavior (or continuum limit)
is observed along the 
transition line that occurs in the small gauge-coupling part of the 
phase diagram.

The paper is organized as follows. In Sec.~\ref{phadiasec} we define 
the compact and the non-compact lattice
AH model and summarize the main features of their phase diagram. In
Sec.~\ref{numres} we define the observables used in the numerical
simulations and present the results of the numerical
analyses. Finally, in Sec.~\ref{conclu} we draw our conclusions.

\section{Compact and noncompact formulations of lattice AH models}
\label{phadiasec}

In this section we define the compact and noncompact formulations of the 
multicomponent lattice AH model on a cubic lattice, 
and summarize the known results for their
phase diagrams. In both formulations the scalar fields are
unit-length $N$-component complex variables ${\bm z}_{\bm x}$ 
associated with the lattice sites. The gauge 
fields are either complex phases 
$\lambda_{{\bm x},\mu}$ (compact model) or real 
numbers $A_{{\bm x},\mu}$ (noncompact model) 
associated with the lattice links.

\subsection{AH model with compact gauge variables}
\label{cogaumod}

In the compact formulation we define 
a gauge variable $\lambda_{{\bm x},\mu}\in{\rm U}(1)$ 
($|\lambda_{{\bm x},\mu}|=1$)
on each lattice link (it starts at site ${\bm x}$ along
one of the lattice direction, $\mu=1,2,3$). The compact AH model
with $N$-component scalar fields of integer charge $q$ is
defined by the partition function
\begin{equation}
Z = \sum_{\{{\bm z},\lambda\}}
e^{-\beta H_c}\,,
\label{copartfun}
\end{equation}
where the Hamiltonian reads
\begin{eqnarray}
H_c &=&  - J N \sum_{{\bm x}, \mu} 2\, {\rm Re}\,(\bar{\bm{z}}_{\bm x} \cdot
\lambda_{{\bm x},\mu}^q \, {\bm z}_{{\bm x}+\hat\mu}) 
\label{chham}\\
&& - \kappa \sum_{{\bm x},\mu>\nu} 2\,{\rm Re}\, (\lambda_{{\bm
      x},{\mu}} \,\lambda_{{\bm x}+\hat{\mu},{\nu}}
  \,\bar{\lambda}_{{\bm x}+\hat{\nu},{\mu}} \,\bar{\lambda}_{{\bm
      x},{\nu}}) \,.\nonumber
\end{eqnarray}
Here the two sums run over all lattice links and plaquettes, respectively.
In the following we rescale $J$ and $\kappa$ by
$\beta$, thus formally setting $\beta=1$. The parameter 
$\kappa\ge 0$ plays the role of inverse gauge coupling.

The compact AH model presents a disordered (confined) phase for small values of
$J$ and one (for $q=1$) or two (for $q\ge 2$) low-temperature ordered
phases for large values of $J$. The transitions between the disordered and
the ordered phases are associated with the breaking of the
global SU($N$) symmetry. The corresponding order parameter is the
gauge-invariant bilinear operator
\begin{equation}
Q^{ab}_{\bm x} = \bar{z}_{\bm x}^a z_{\bm x}^b - {1\over N} \delta^{ab}\,.
\label{qdef}
\end{equation}
For $\kappa=0$ the model is equivalent to a particular
lattice formulation of the CP$^{N-1}$ model, which undergoes a phase
transition at a finite value of $J$ (see, e.g.,  Ref.~\cite{PV-19-CP}).
In the $\kappa\to\infty$ limit the model reduces to an O($2N$)
vector model, which presents a transition at a finite value of $J$, as
well. 

\begin{figure}[tbp]
\includegraphics*[width=0.95\columnwidth]{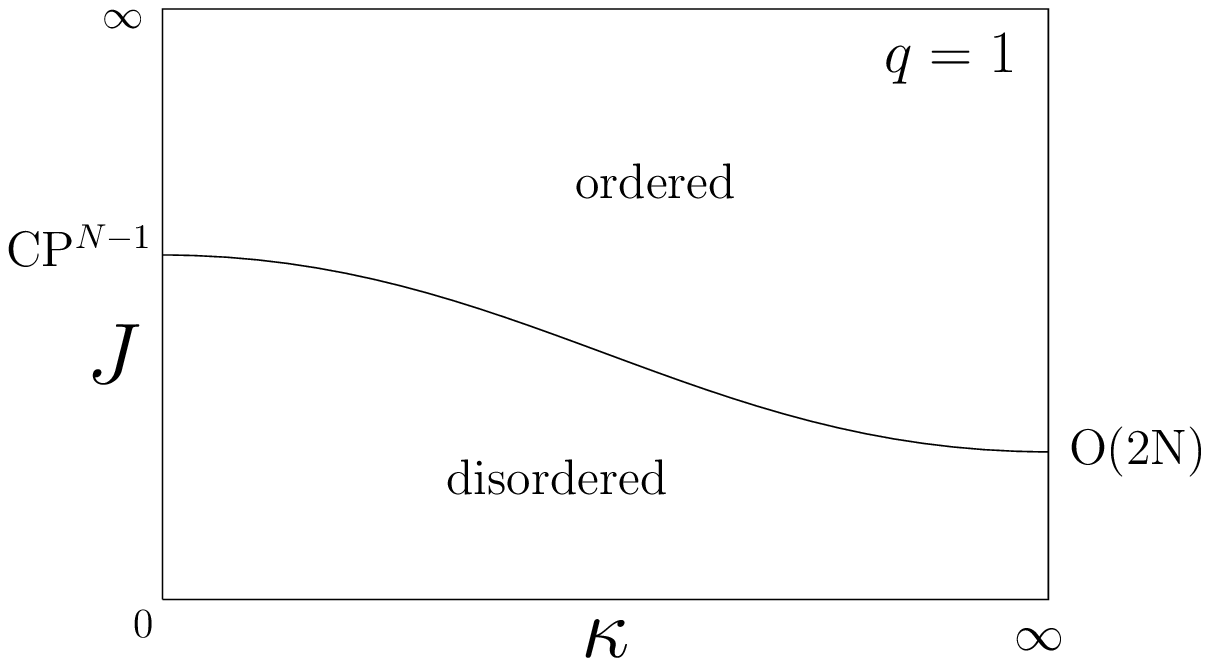}
\includegraphics*[width=0.95\columnwidth]{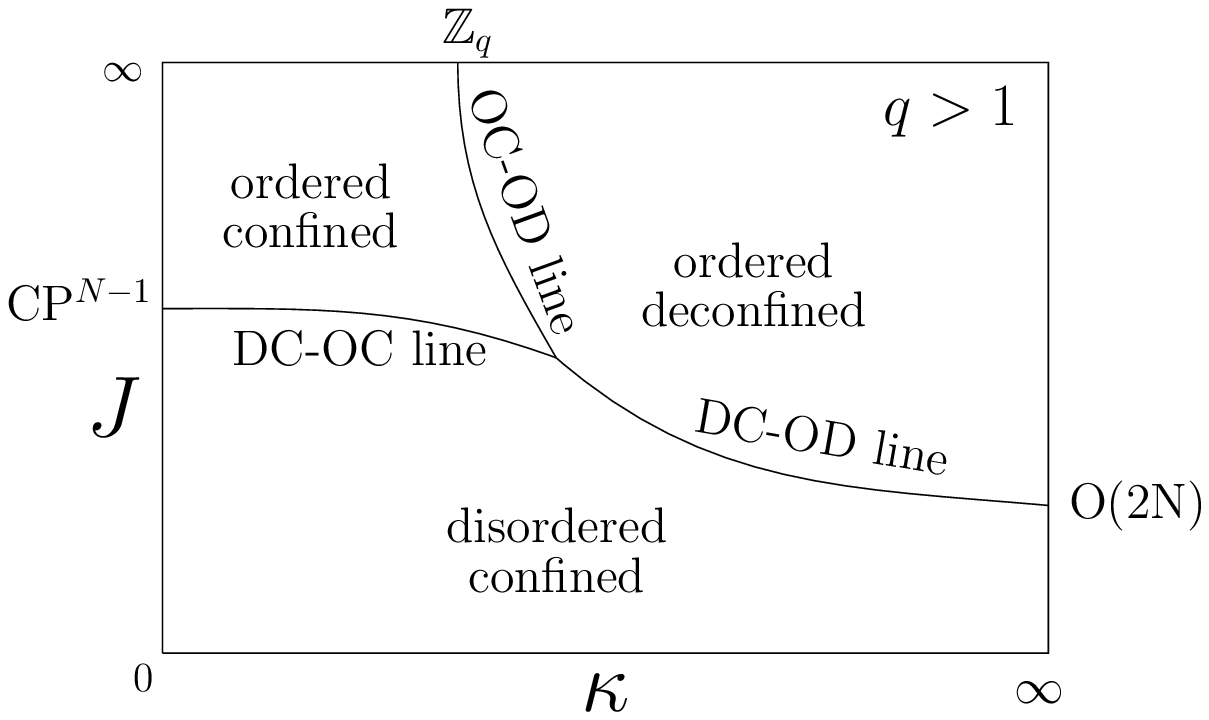}
  \caption{Sketch of the phase diagram of the 3D
    compact lattice AH model, in which a compact
    U(1) gauge field is coupled to an $N$-component unit-length
    complex scalar field with charge $q$, for generic $N\ge
    2$. In the upper panel 
    we report the phase diagram for $q=1$, with two phases
    separated by a single transition line. In the lower panel, we report the 
    phase diagram for $q=2$, with three phases, the disordered-confined (DC),
    the ordered-deconfined (OD), and the ordered-confined (OC) phases.
    The AH model is equivalent to the CP$^{N-1}$ model for
    $\kappa=0$, to the O($2N$) vector model for $\kappa\to\infty$. 
    For $J\to\infty$ and $q\ge 2$, we obtain 
    the lattice ${\mathbb Z}_q$ gauge model.}
\label{phdiasketchqLAH}
\end{figure}

For $q=1$, only two phases are present, see Fig.~\ref{phdiasketchqLAH} (top):
a disordered phase for small $J$ and an
ordered phase for large $J$. They are separated by a single transition line, 
along which only gauge-invariant scalar modes become critical.
Gauge fields do not develop long-range correlations,
but they prevent gauge-dependent scalar correlations, such as the
vector correlations $\langle \bar{\bm z}_{\bm x}\cdot {\bm z}_{\bm
  y}\rangle$, from becoming critical. As a consequence, the
critical behavior is described by a LGW
$\Phi^4$ theory in terms of a gauge-invariant scalar order parameter.
The fundamental field  is a traceless hermitian matrix field
$\Psi^{ab}({\bm x})$,
which can be formally defined by coarse graining the lattice order
parameter $Q_{\bm x}^{ab}$, defined in Eq.~(\ref{qdef}).
The LGW field theory is obtained by
considering the most general fourth-order polynomial in $\Psi$
consistent with the U($N$) global symmetry~\cite{PTV-17,PV-19-CP}:
\begin{eqnarray}
{\cal L}_{\rm LGW} &=& {\rm Tr} (\partial_\mu \Psi)^2 
+ r \,{\rm Tr} \,\Psi^2 \label{hlg}\\
&+&   w \,{\rm tr} \,\Psi^3 
+  \,u\, ({\rm Tr} \,\Psi^2)^2  + v\, {\rm Tr}\, \Psi^4\, .
\nonumber
\end{eqnarray}
In this approach, continuous
transitions are possible only if the RG flow in the LGW theory
has a stable fixed point.
For $N=2$ the Lagrangian (\ref{hlg}) is
equivalent to that of the O(3) vector model (in particular, the
$\Psi^3$ term cancels), thus continuous transitions in the
Heisenberg universality class~\cite{PV-02} can be observed in the 
$N=2$ AH model. For larger values of $N$, 
the LGW approach predicts all transitions to be of first order, 
because of the presence of the
$\Psi^3$ term~\cite{PV-19-AH3d,PV-19-CP,PV-20-largeNCP}.

For $q\ge 2$ the phase diagram is more complex, see
Fig.~\ref{phdiasketchqLAH} (bottom), with three different phases
~\cite{FS-79, SSSNH-02, SSNHS-03, NSSS-04, CFIS-05, CIS-06, WBJS-08, 
BPV-20-hcAH}.  
They are characterized by the large-distance
behavior of both scalar and gauge observables. Beside the 
scalar gauge-invariant observable (\ref{qdef}), one may consider
the Wilson loop of the gauge fields,
that signals the confinement or deconfinement of charge-one
external static sources.  As shown in Fig.~\ref{phdiasketchqLAH}, for
small $J$ and any $\kappa \ge 0$, there is a phase in which
scalar-field correlations are disordered and single-charge particles
are confined (the Wilson loop obeys the area law).  For large values
of $J$ (low-temperature region) scalar correlations are ordered and
the SU($N$) symmetry is broken. Two different phases occur here: for
small $\kappa$, single-charge particles are confined, while they are
deconfined for large $\kappa$.

The three different phases are separated by three transition lines
meeting at a multicritical point: the DC-OD transition line between
the disordered-confined (DC) and the ordered-deconfined (OD) phases,
the DC-OC line between the disordered-confined and ordered-confined
(OC) phases, and the OC-OD line between the ordered-confined and
ordered-deconfined phases.  The transition lines have different
features, since they are associated with different phases.  Moreover,
their nature depends on the number $N$ of components and on the charge
$q$ of the scalar matter.  The transitions along the DC-OC line are
the same as that in the 3D CP$^{N-1}$ model for $\kappa=0$. They
are continuous for $N=2$, belonging to the O(3) vector universality
class, and of first order for $N\ge 3$. For $J=\infty$, 
the model (\ref{chham}) is equivalent to a ${\mathbb Z}_q$ gauge
model~\cite{BPV-20-hcAH}.
A natural hypothesis is that the 
transitions along the OC-OD line belong to the universality 
class of the ${\mathbb Z}_q$ gauge model.
This hypothesis has been verified numerically for
$q=2$~\cite{BPV-20-hcAH}, for which 
$\kappa_c = 0.380706646(6)$ in the
limit $J\to\infty$. Finally, transitions along the DC-OD line are
continuous for large values of $N$, as we
shall see below, and belong to the same universality class for any
$q\ge 2$. We shall argue that they realize the continuum limit of
the AH field theory (\ref{AHFT}).

\subsection{AH model with noncompact gauge variables}
\label{nocogaumod}

In the noncompact formulation the fundamental gauge variable is the real
vector field $A_{{\bm x},\mu}$. The lattice Hamiltonian reads
\begin{eqnarray}
H_{nc} &=& - J N \sum_{{\bm x}, \mu} 2\, {\rm Re}\,(\bar{\bm{z}}_{\bm
  x} \cdot e^{iA_{{\bm x},\mu}} \, {\bm z}_{{\bm x}+\hat\mu})
\label{nchham}\\
&& + {\kappa_g\over 2} \sum_{{\bm x},\mu>\nu} 
(\Delta_{\hat\mu} A_{{\bm x},\nu} - 
\Delta_{\hat\nu} A_{{\bm x},\mu})^2\,,
\nonumber
\end{eqnarray}
where the sums run
over all links and plaquettes, respectively, 
$\Delta_{\hat\mu} A_{\bm x} \equiv
A_{{\bm x}+\hat\mu} - A_{\bm x}$, and $\kappa_g\ge 0$ corresponds to
the inverse gauge coupling~$1/g^2$ of the continuum theory
(\ref{AHFT}).  The partition function reads
\begin{equation}
Z_{nc} = \sum_{\{{\bm z},A\}} e^{-H_{nc}}\,.
\label{ncz}
\end{equation}
Unlike the compact case, the charge $q$ of the scalar field is
irrelevant: We can set $q=1$ by
a redefinition of the gauge field $A_{\bm x}$.

At variance with the compact case, the partition function $(\ref{ncz})$ is only
formally defined. Since the integration domain for the gauge variables is
noncompact, gauge invariance implies $Z_{nc} = \infty$ even on a finite
lattice. If periodic boundary conditions are used, this problem is present even
when  a maximal gauge fixing is added. Indeed, the partition function still
diverges because of the presence of gauge-invariant zero modes: noncompact
gauge-invariant Polyakov operators, i.e., sums of the fields $A_{{\bm x},\mu}$
along nontrivial paths winding around the lattice \cite{BPV-21-ncAH}, are still
unbounded.  To overcome this problem, $C^*$ boundary conditions~\cite{KW-91,
LPRT-16} were considered in Ref.~\cite{BPV-21-ncAH}. These boundary conditions
preserve gauge invariance and provide a rigorous definition of the partition
function in a finite volume.

\begin{figure}[tbp]
\includegraphics*[width=0.95\columnwidth]{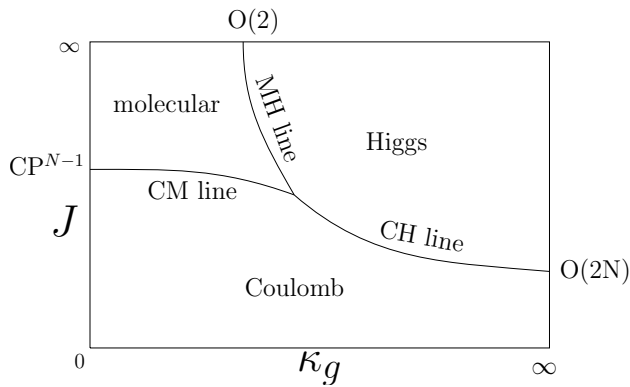}
  \caption{Sketch of the phase diagram of the lattice
    AH model with noncompact gauge fields and unit-length
    $N$-component complex scalar fields, for generic $N\ge 2$. There
    are three different phases, the Coulomb, Higgs and molecular
    phases, and three transition lines: the Coulomb-to-Higgs (CH)
    line between the Coulomb and Higgs phases, the
    Coulomb-to-molecular (CM) line, and the molecular-to-Higgs (MH)
    line. The model is equivalent to 
    the CP$^{N-1}$ model for $\kappa_g=0$, to the O($2N$) vector
    model for $\kappa_g\to\infty$, and to the inverted $XY$ model
    for $J\to\infty$.  }
\label{phdiasketchncLAH}
\end{figure}

In Fig.~\ref{phdiasketchncLAH} we sketch the phase diagram of the
noncompact lattice AH model.
For any $N\ge 2$ the phase diagram is characterized by three
phases. For small $J$ we have a Coulomb phase, 
in which the global SU($N$) symmetry
is unbroken and electromagnetic correlations are-long ranged. For large $J$,
there are two phases characterized by the breaking of the SU($N$)
symmetry. They are distinguished by the behavior of the gauge
modes. In the Higgs phase (large $\kappa$), 
electromagnetic correlations are gapped,
while in the molecular phase (small $\kappa$) 
the electromagnetic field is ungapped.

The Coulomb, molecular, and Higgs phases are separated by three
different transition lines meeting at a multicritical point:
the CM line between the Coulomb and molecular phases, the MH line
between the molecular and Higgs phases, and the CH line between the
Coulomb and Higgs phases.  Their nature crucially depends on the
number $N$ of components.  The transitions along the CM line are
the same as that in the 3D CP$^{N-1}$ model ($\kappa_g=0$): 
they are continuous for $N=2$, belonging to the O(3) vector
universality class, and of first order for $N\ge 3$. The transitions
along the MH line are expected to be continuous, and to belong to the
$XY$ universality class, at least for sufficiently large values of the
parameter $J$ [the transition point in the limit $J\to\infty$ is
located at $\kappa_{gc} = 0.076051(2)$, obtained by using the
estimate $\beta_c=3.00239(6)$ reported in Ref.~\cite{NRR-03} and
identifying $\kappa_c = \beta_c/(4\pi^2)$]. Finally, transitions along
the CH line are continuous for a sufficiently large number $N$ of
components. As argued in Ref.~\cite{BPV-21-ncAH}, they should realize
the continuum limit of the AH field theory (\ref{AHFT}).

\subsection{Relation between the compact and the noncompact model}
\label{qinfty}

It is interesting to note that the compact
formulation is equivalent to the noncompact one for $q\to\infty$.
Indeed, if we rewrite the compact field $\lambda_{{\bm x},\mu}$ as 
\begin{equation}
    \lambda_{{\bm x},\mu} = e^{i A_{{\bm x},\mu}/q}
\end{equation}
with $A_{{\bm x},\mu} \in [- \pi q, \pi q]$, the 
Hamiltonian (\ref{chham}) becomes 
\begin{eqnarray}
H_{c} &=& - J N \sum_{{\bm x}, \mu} 2\, {\rm Re}\,(\bar{\bm{z}}_{\bm
  x} \cdot e^{iA_{{\bm x},\mu}} \, {\bm z}_{{\bm x}+\hat\mu})
\label{chham2}\\
&& - 2 {\kappa} \sum_{{\bm x},\mu>\nu}  
\hbox{Re } \exp\left[-{i\over q} (\Delta_{\hat\mu} A_{{\bm x},\nu} -
\Delta_{\hat\nu} A_{{\bm x},\mu}) \right].
\nonumber
\end{eqnarray}
For $q\to \infty$, the gauge fields $A_{{\bm x},\mu}$ become unbounded 
and the Hamiltonian
is equivalent to that of the noncompact formulation,
provided that $\kappa_g=2\kappa/q^2$. Note that the equivalence 
trivially holds as long as the
fluctuations of $A_{{\bm x},\mu}$ on each plaquette are bounded and 
uncorrelated for $q\to \infty$, i.e., for any point of the phase diagram
except possibly at phase transitions.
Therefore, the
noncompact formulation (\ref{ncz}) should be recovered from the
compact formulation (\ref{chham}) in the limit $q\to\infty$,
keeping $\kappa_g=2\kappa/q^2$ fixed.

The equivalence of the models also holds for $J\to \infty$.
In this limit the compact formulation reduces to
the $\mathbb{Z}_{q}$ model
\begin{equation}
H_q=-\kappa^{(q)}\sum_{{\bm x},\mu>\nu}
{\rm Re}\, (\lambda_{{\bm
      x},{\mu}} \,\lambda_{{\bm x}+\hat{\mu},{\nu}}
  \,\bar{\lambda}_{{\bm x}+\hat{\nu},{\mu}} \,\bar{\lambda}_{{\bm
      x},{\nu}}) \,,
\end{equation}
where $\kappa^{(q)}=2\kappa$ and the gauge field takes 
the values $\lambda_{{\bm x},\mu}=e^{i\frac{2\pi}{q}n}$, with $n\in
0,1,\ldots,q-1$. If the limit $q\to\infty$ is smooth, 
the critical value of the coupling $\kappa^{(q)}$ should scale as
\begin{equation}
\kappa^{(q)}_c\simeq \kappa_{gc} \, q^2
\end{equation} 
for large $q$, where $\kappa_{gc}=0.076051(2)$ is the critical
coupling of the inverted XY model that represents the $J\to\infty$ limit
of the noncompact model \cite{NRR-03}.
The $q$-dependence of $\kappa^{(q)}_c$ has been
numerically investigated in Refs.~\cite{Bhanot:1980pc,Borisenko:2013xna}.
Ref.~\cite{Borisenko:2013xna} determined the large-$q$ behavior,
obtaining
\begin{equation}
\kappa^{(q)}_c\simeq C q^2,
\end{equation} 
with $C=0.076053(4)$ [we use the estimate $A=1.50122(7)$
reported in Ref.~\cite{Borisenko:2013xna}, identifying
$C=A/(2\pi^2)$], which is in excellent agreement with the 
estimate of $\kappa_{gc}$.

The argument presented above only proves that the compact model converges 
to the noncompact one as $q\to \infty$, but does not provide us with 
any information on the critical behavior. For the ${\mathbb Z}_q$ transition
observed for $J\to \infty$, numerical results \cite{Borisenko:2013xna} 
indicate that the transition belongs to the XY universality class for any 
$q \ge 5$. Thus, for these values of $q$, the compact 
${\mathbb Z}_q$ model and the noncompact inverted XY model have a transition
in the same universality class. 
In the next section we will present numerical results showing that
the same occurs at the transitions controlled by the CFP of the AH field
theory. For $N$ large enough and any $q\ge 2$, the transitions
along the CH line of the noncompact model and along the
DC-OD line of the compact model belong to 
the same universality class, controlled by the CFP.

\section{Numerical analyses}
\label{numres}

We have performed MC simulations of the compact AH model with 
$N=15$ and $N=25$ and some values of $q\ge 2$. We use $C^*$ boundary 
conditions, as we did for the noncompact model \cite{BPV-21-ncAH}.
This allows us to compare the FSS results---universal 
scaling curves depend on boundary conditions---for the 
compact model with those for the noncompact one. 
The results of the FSS analyses of the MC data will provide  
strong evidence that, 
for any $q\ge 2$, the continuous transitions
along the DC-OD transition line, running up to $\kappa\to\infty$,
belong to the same universality class as those along the 
CH transition line of the noncompact formulation.

We will also compute the correlation-length exponent $\nu$ 
and the exponent $\eta_q$ that characterizes the singular behavior of the 
susceptibility of the bilinear field $Q_{\bm x}$. The results will
be compared with the large-$N$ predictions \cite{HLM-74,YKK-96}
\begin{eqnarray}
\nu &=& 1 - \frac{48}{\pi^2 N} + O(N^{-2})\,, 
\label{nulargen} \\
\eta_q &=& 1 - \frac{32}{\pi^2 N}  + O(N^{-2})\, .
\label{etalargen}
\end{eqnarray} 
The good agreement of the numerical estimates of the
critical exponents with the large-$N$ field-theory expressions
demonstrates that these continuous transitions are associated with the CFP
of the AH field theory (\ref{AHFT}).

\subsection{Observables and finite-size scaling}
\label{observables}

To characterize phase transitions associated with the breaking of the
SU($N$) symmetry, we consider correlations of the gauge-invariant
bilinear operator $Q$ defined in Eq.~(\ref{qdef}). Since
$Q$ is periodic when using $C^*$ boundary conditions, its two-point
correlation function can be defined as
\begin{equation}
G({\bm x}-{\bm y}) = \langle {\rm Tr}\, Q_{\bm x} Q_{\bm y} \rangle\,.
\label{gxyp}
\end{equation}
The corresponding susceptibility and correlation length are defined as
$\chi=\sum_{\bm x} G({\bm x})$ and
\begin{eqnarray}
\xi^2 \equiv  {1\over 4 \sin^2 (\pi/L)}
{\widetilde{G}({\bm 0}) - \widetilde{G}({\bm p}_m)\over 
\widetilde{G}({\bm p}_m)}\,,
\label{xidefpb}
\end{eqnarray}
where $\widetilde{G}({\bm p})=\sum_{{\bm x}} e^{i{\bm p}\cdot {\bm x}}
G({\bm x})$ is the Fourier transform of $G({\bm x})$, and ${\bm p}_m =
(2\pi/L,0,0)$. 

In our analysis we consider RG invariant quantities, such as $R_\xi =
\xi/L$ and the Binder parameter
\begin{equation}
U = {\langle \mu_2^2\rangle \over \langle \mu_2 \rangle^2} \,, \qquad
\mu_2 = 
\sum_{{\bm x},{\bm y}} {\rm Tr}\,Q_{\bm x} Q_{\bm y}\,.
\label{binderdef}
\end{equation}
At a continuous phase transition, any RG invariant ratio $R$, scales
as~\cite{PV-02}
\begin{eqnarray}
R(j,L) = f_R(X) +  L^{-\omega} g_R(X) + \ldots \,, \label{scalbeh}
\end{eqnarray}
where 
\begin{equation}
  X = (j-j_c)L^{1/\nu}\, .
  \label{Xdef}
\end{equation}
Here $\nu$ is the correlation-length critical exponent, 
$\omega$ is the leading correction-to-scaling exponent, 
$j$ is the Hamiltonian parameter driving the transition, and 
$j_c$ is the critical point (we will perform simulations varying
$J$ at fixed $\kappa$ or $\kappa_g$ for the compact and noncompact
model, respectively, so that $j$ should be identified with $J$).
The function $f_R(X)$ is universal up to a multiplicative rescaling of its
argument. 
Assuming that $R_\xi$ is a monotonically increasing 
function of $j$, we can combine the RG predictions for $U$ and $R_\xi$ to
obtain
\begin{equation}
  U(j,L) = F(R_\xi) + O(L^{-\omega})\,,
\label{uvsrxi}
\end{equation}
where $F$ depends only on the universality class, boundary conditions,
and lattice shape, without nonuniversal factors.
Eq.~(\ref{uvsrxi}) is particularly convenient because it allows us to
test universality-class predictions without requiring a tuning of
nonuniversal parameters.

The exponent $\nu$ will be determined from the FSS behavior of $R_\xi$ and 
$U$, assuming the scaling behavior (\ref{scalbeh}). The exponent
$\eta_q$ will be computed from the scaling behavior of the susceptibility
$\chi$. In the FSS limit, it scales as 
\begin{equation}
\chi(j,L) = L^{2-\eta_q} \left[f_\chi(X) + L^{-\omega} g_\chi(X)+\ldots
\right],
\end{equation}
where $X$ is defined in Eq.~(\ref{Xdef}).

\subsection{Monte Carlo results}
\label{MCres}

\begin{figure}[tbp]
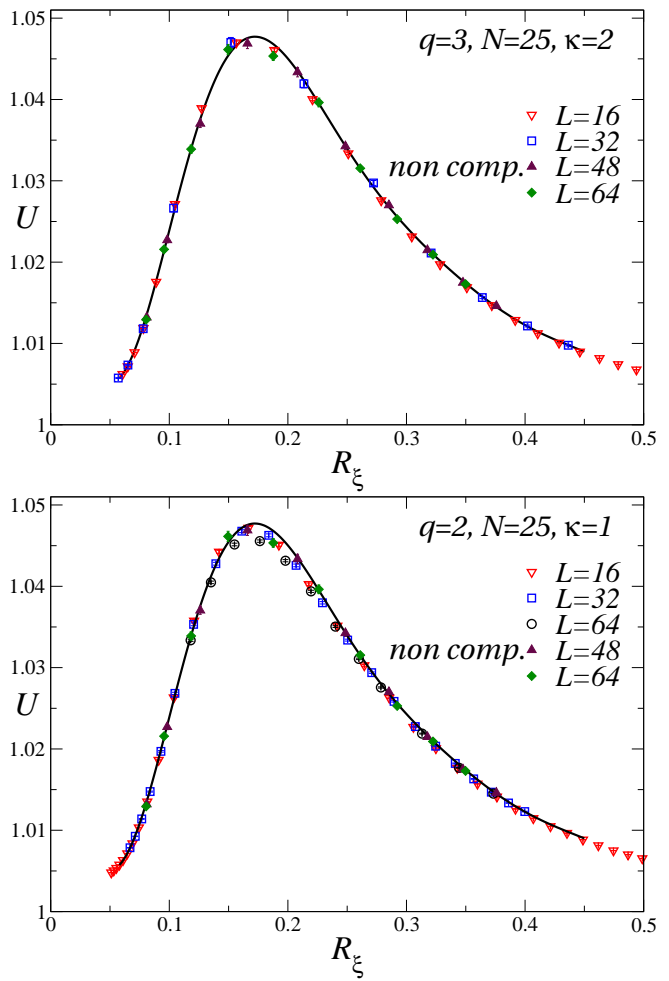

  \includegraphics*[width=1.0\columnwidth]{urxi-q3-n25.eps}
  \includegraphics*[width=1.0\columnwidth]{urxi-q2-n25.eps}
  \caption{Estimates of $U$ versus $R_\xi$ for the compact AH model
  with $N=25$ and $C^*$ boundary conditions. 
  Top: results for $q=3$, $\kappa=2$; bottom: results for $q=2$, $\kappa=1$.
  The continuous line in each panel is an extrapolation of data for 
  the noncompact AH model~\cite{BPV-21-ncAH}; 
  noncompact-model data for $L=48$ and $L=64$ are also
  reported to provide an estimate of the accuracy of the extrapolation.
  }
  \label{n25q23}
\end{figure}

Let us first report our results for $N=25$. We have considered two values of 
$q$, $q=2$ and 3, and, for each of them, we have performed simulations at 
a fixed value of $\kappa$, chosen so that the transition belongs to the 
DC-OD line. For $q=2$, simulations were already performed \cite{BPV-20-hcAH}
fixing $\kappa = 1$ and 
using periodic boundary conditions, identifying the transition at 
$J_c = 0.29333(3)$. 
For $q=3$, the results of Ref.~\cite{Borisenko:2013xna} indicate that 
the OC-OD line ends at $\kappa_c=0.5422(1)$ for $J\to\infty$. To be on the 
safe side, 
we have performed simulation keeping $\kappa = 2$ fixed, 
observing a transition for 
$J=J_c\approx 0.2945$.

To verify whether the transitions for $q=2$ and $q=3$ belong to the 
same universality class as the transitions in the noncompact model along the 
CH line, in Fig.~\ref{n25q23} we report the Binder
parameter $U$ versus the ratio $R_\xi$. The compact-model data fall on top 
of the curve obtained from simulations of the noncompact
model~\cite{BPV-21-ncAH}.  The agreement is excellent for both values of $q$.
These results
demonstrate that the continuous transitions in the compact model 
(DC-OD line) and in the noncompact model (CH line) all 
belong to the same universality class.

For $q=2$ we also estimated the critical exponents, performing the same 
analysis we did in Refs.~\cite{BPV-20-hcAH,BPV-21-ncAH}. 
To estimate the exponent 
$\nu$, we perform combined fits of $U$ and $R_\xi$ to 
Eq.~(\ref{scalbeh}).  We parametrize the scaling functions $f_R(X)$ and 
$g_R(X)$ with polynomials (we use 24th-order and 8th-order polynomials for
the two functions, respectively). We perform fits including only data 
with $L\ge 16$, varying the exponent $\omega$ in the range $[0.6,1.2]$ 
(results depend marginally on the value of this exponent). We only
consider data in the interval $X\in [X_{\rm min},X_{\rm max}]$, varying 
$X_{\rm min}$ (between $-0.5$ and $-0.3$) and $X_{\rm max}$ (between
0.15 and 0.25). Results are stable. We obtain 
$J_c=0.293331(2)$ (in excellent agreement with the estimate
of Ref.~\cite{BPV-20-hcAH} reported above) and 
\begin{equation}
\nu=0.817(7).
\label{nuest-N25}
\end{equation}
The error includes the statistical error and also takes 
into account the variation
of the estimate as the fit parameters are changed.

To estimate $\eta_q$, we have performed fits to
\begin{equation}
\ln \chi = (2 - \eta_q) \ln L + h_{1\chi}(X) +  L^{-\omega} h_{2\chi}(X),
\end{equation}
parametrizing $h_{1\chi}(X)$ and $h_{2\chi}(X)$ with polynomials. 
In this case fits are sensitive to the value of $\omega$. We end up with 
$\omega = 1.05(10)$ and 
\begin{equation}
\eta_q = 0.882(2).
\label{etaest-N25}
\end{equation}
The estimates (\ref{nuest-N25}) and (\ref{etaest-N25}) are 
significantly more accurate than, but consistent with 
previous determinations. Ref.~\cite{BPV-21-ncAH} obtained $\nu=0.802(8)$
for the compact model with $q=2$, while 
Ref.~\cite{BPV-20-hcAH} reported $\nu=0.815(15)$ for the noncompact model.
As for $\eta_q$, previous estimates are 
$\eta_q=0.88(2)$ (compact model with $q=2$),
and $\eta_q=0.883(7)$ (noncompact model).

\begin{figure}[tbp]
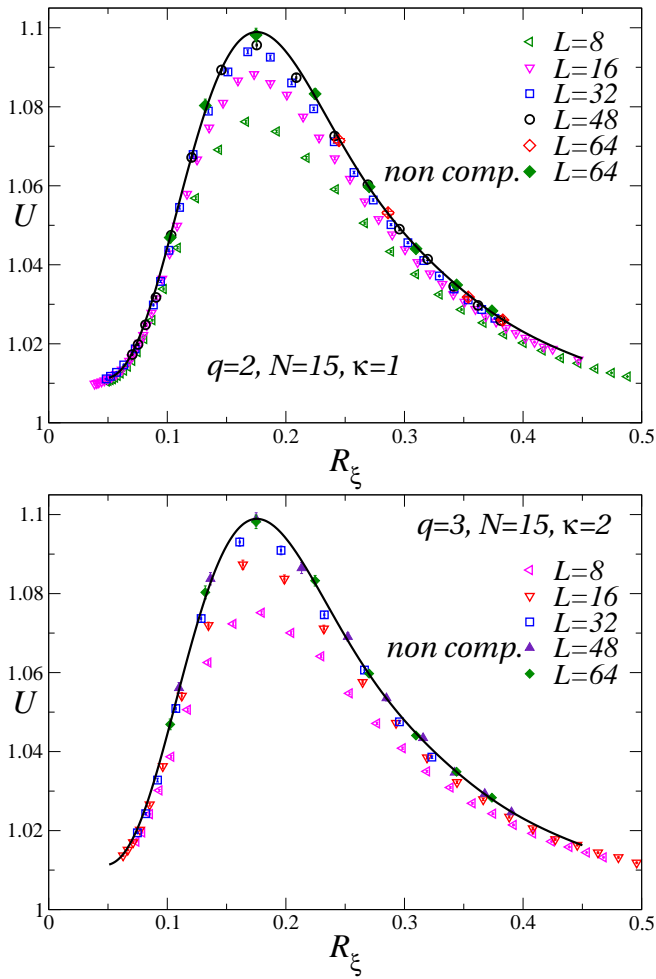

  \includegraphics*[width=1.0\columnwidth]{urxi-q2-n15.eps}
  \includegraphics*[width=1.0\columnwidth]{urxi-q3-n15.eps}
  \caption{Estimates of $U$ versus $R_\xi$ for the compact AH model with $N=15$
  and $C^*$ boundary conditions. Top: results for $q=2$ and $\kappa=1$;
  bottom: results for $q=3$ and $\kappa=2$.
  The continuous line is an extrapolation of  the data for 
  the noncompact model~\cite{BPV-21-ncAH};
  noncompact-model data for $L=48$ and $L=64$ are also
  reported to provide an estimate of the accuracy of the extrapolation.
  }
  \label{n15q23}
\end{figure}

\begin{figure}[tbp]
  \includegraphics*[width=1.0\columnwidth]{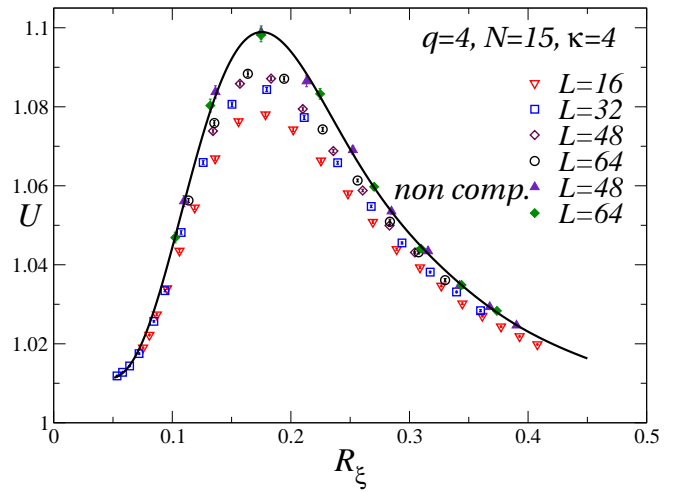}
  \caption{Estimates of $U$ versus $R_\xi$ for the compact AH model with $N=15$
  and $C^*$ boundary conditions. Results for $q=4$ and $\kappa=4$;
  The continuous line is an extrapolation of  the data for 
  the noncompact model~\cite{BPV-21-ncAH};
  noncompact-model data for $L=48$ and $L=64$ are also
  reported to provide an estimate of the accuracy of the extrapolation.
  }
  \label{n15q4}
\end{figure}

We have performed a similar analysis for $N=15$. In this case
we have considered $q = 2$, $q=3$, and $q=4$, performing simulations at fixed
$\kappa$ along the DC-OD line. For $q=2$ and $q=3$ we have 
performed simulations at $\kappa =1$ (transition at
$J\approx 0.307$) and at $\kappa = 2$ (transition at
$J\approx 0.308$), respectively, as we did for $N=25$. For $q=4$, we have
chosen $\kappa = 4$. Given that the OC-OD line ends at 
\cite{Bhanot:1980pc, Borisenko:2013xna}, $\kappa = 0.76135(2)$, 
$J=\infty$, this choice should guarantee that the transition we observe 
for $J\approx 0.304$ belongs to the DC-OD line.

The results for the Binder parameter as a function of $R_{\xi}$ are
reported in Fig.~\ref{n15q23} for $q=2$ and $q=3$. Once again, data for the
noncompact lattice AH model with $N=15$ (from Ref.~\cite{BPV-21-ncAH}) are also
shown for comparison.  In this case scaling corrections are larger than for
$N=25$. Nonetheless, data approach the expected scaling curve
when increasing the lattice size.  Results for $q=4$ 
are shown in Fig.~\ref{n15q4}. In this case, corrections to scaling are 
large and, in spite of the large lattices considered---we performed
simulations up to $L=64$---the compact-model data are not yet close to the 
noncompact-model curve, although they show the correct trend as $L$ increases.
Most probably, this is a crossover effect due to the O($2N$) fixed
point that controls the critical behavior for  $\kappa=\infty$. Indeed, 
its presence gives rise to 
crossover effects that increase with $\kappa$ and that can become particularly
strong for the simulations that have been performed along the line
with $\kappa = 4$.

As we did for $N=25$, we also compute the critical exponents.  We consider only
the data with $q=2$, since scaling corrections appear to be smaller than for
$q=4$ and only a limited number of lattice sizes is available $q=3$.  The combined analysis of
the Binder parameter and of $R_\xi$ gives $J_c = 0.306957(4)$ and 
\begin{equation}
\nu = 0.728(5),
\label{nuest-N15}
\end{equation}
which is in agreement with the noncompact-model 
estimate $\nu = 0.721(3)$, obtained in
Ref.~\cite{BPV-21-ncAH}. Again the error takes into account statistical errors
and how the estimate changes as the fit parameters are varied. In particular,
the quoted result is consistent with an exponent $\omega$ varying between
0.6 and 1. 
We also estimate the exponent $\eta_q$, obtaining 
\begin{equation}
\eta_q = 0.815(3),
\label{etaest-N15}
\end{equation}
which is in full agreement with the estimate 0.815(10), obtained in 
the noncompact model \cite{BPV-21-ncAH}.

\begin{figure}[tbp]
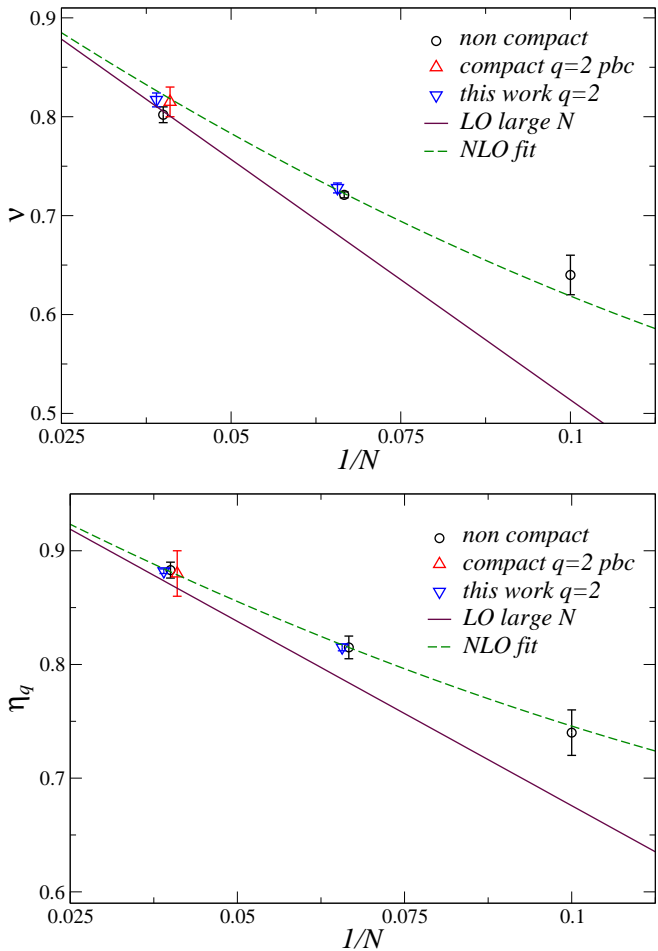

  \includegraphics*[width=1.0\columnwidth]{crit_exp_nu.eps}\\
  \vspace{0.25cm}
  \includegraphics*[width=1.0\columnwidth]{crit_exp_eta.eps}
  \caption{Critical exponents $\nu$ (top) and $\eta_q$ (bottom) 
  versus $1/N$. We report:
  results ($N=10,15,25$) for the noncompact model \cite{BPV-21-ncAH};
  results ($N=25$ only) 
  for the compact model with $q=2$ and periodic boundary 
  conditions (pbc)~\cite{BPV-20-hcAH}; the results for $N=15$ and 25
  of the present work ($q=2$); the leading-order (LO) large-$N$ predictions 
  [Eqs. (\ref{nulargen}) and (\ref{etalargen})]; the phenomenological
  next-to-leading (NLO) interpolations, Eq.~(\ref{NLO-largeN}) with 
  $a_\nu = 10.5$, $a_\eta = 7.0$.
  }
  \label{critexpfig}
\end{figure}

The estimates of the critical exponents for $N=15$ and 25 are displayed in
Fig.~\ref{critexpfig}, together with the leading-order large-$N$ estimates
Eqs.~(\ref{nulargen}) and (\ref{etalargen}). They would predict $\nu = 0.805$,
$\eta_q = 0.870$ for $N=25$ and $\nu = 0.676$, $\eta_q = 0.784$ for $N=15$, to
be compared with the previosuly obtained numerical results $\nu=0.817(7)$,
$\eta_q=0.882(2)$ for $N=25$ [see Eqs.~(\ref{nuest-N25}) and (\ref{etaest-N25})], and
$\nu=0.728(5)$, $\eta_q=0.815(3)$ for $N=15$ [see Eqs.~(\ref{nuest-N15}) and
(\ref{etaest-N15})].  The values of the critical exponents for $N=25$ are very
close to their leading-order large-$N$ estimates, and deviations from the
$O(N^{-1})$ asymptotic behavior are consistent with a next-to-leading
$O(N^{-2})$ correction.  If we assume 
\begin{equation}
\begin{aligned}
\nu = 1 - \frac{48}{\pi^2 N} + \frac{a_\nu}{N^2}\,,\\
\eta_q = 1 - \frac{32}{\pi^2 N}  + \frac{a_\eta}{N^2}\,.
\end{aligned}
\label{NLO-largeN}
\end{equation}
and we fix the unknown parameters by requiring these expressions to be exact
for $N=25$, we obtain the estimates $a_\nu = 7(4)$ and $a_\eta = 7(1)$. Using
these values, we would predict $\nu = 0.708(19)$  and $\eta_q = 0.816(6)$ for
$N=15$, in agreement with the estimates (\ref{nuest-N15}) and
(\ref{etaest-N15}).  For $N=10$ we would predict $\nu = 0.59(4)$ and $\eta_q =
0.749(13)$, again in substantial agreement with the results $\nu = 0.64(2)$,
$\eta_q = 0.74(2)$ of Ref.~\cite{BPV-21-ncAH} for the noncompact model. By
fitting to Eq.~(\ref{NLO-largeN}) all the results for $\nu$ and $\eta_q$
obtained in this work, in the noncompact model \cite{BPV-21-ncAH} and in
compact model with periodic boundary conditions (only $q=2$, $N=25$)
\cite{BPV-20-hcAH} we obtain 
\begin{equation}
a_\nu = 10.5(5) \qquad a_\eta = 7.0(5)\ ,
\end{equation}
and the results of this phenomenological interpolation
are shown in Fig.~\ref{critexpfig}.

\section{Conclusions}
\label{conclu}

We have investigated whether and under which conditions the 3D multicomponent
AH field theory (scalar electrodynamics) is realized as the continuum limit of
statistical lattice gauge models.  For this purpose we consider a lattice model
with unit-length degenerate $N$-component scalar fields of charge $q$ coupled
to compact gauge fields with U(1) local and SU($N$) global invariance. 

The FSS analyses of the MC results show that, for $q\ge 2$,  the transitions
along the  line that separates the confined and deconfined phases, see
Fig.~\ref{phdiasketchqLAH} (bottom), are continuous for a sufficiently large
number of components (we perform a detailed study for $N=15$ and 25) and that
they belong to the same universality class for any $q\ge 2$. Morover, they are
in the same universality class as the transitions along the CH line (see
Fig.~\ref{phdiasketchncLAH}), in the lattice AH model with noncompact gauge
fields.  Since both scalar and gauge correlations are critical along the CH
line, the  effective field-theory description of these transitions is provided
by the AH field theory with Lagrangian (\ref{AHFT}), with explicit gauge
fields.  The stable CFP point of the RG flow of the AH field theory, which is
present for $N\ge N_3^*$ with $N_3^*=7(2)$, should characterize the universal
features of these transition lines (the OC-OD line in the compact model with
$q\ge 2$,  see Fig.~\ref{phdiasketchqLAH} (bottom), and the CH line in the noncompact
model, see Fig.~\ref{phdiasketchncLAH}).

We believe that these results improve our understanding of the critical
behavior (continuum limit) of gauge field theories in dimension lower than
four, which are relevant in condensed-matter physics, see, e.g.,
Refs.~\cite{Sachdev-19,Anderson-book,Wen-book}.  In particular, they shed light
on the conditions under which we may expect to observe transitions controlled
by the CFP of the RG flow of 3D gauge field theories. This issue is also
relevant for  nonabelian gauge theories with matter fields; see, e.g.,
Refs.~\cite{BFPV-21-sunfu,BFPV-21-sunadj,SSST-19,SPSS-20} for related
discussions. We believe that further investigations are called for, to achieve
a satisfactory understanding of the nonperturbative regimes of abelian and
nonabelian gauge field theories.

\medskip

\emph{Acknowledgement}. Numerical simulations have been performed using the
Green Data Center of the University of Pisa.

\end{document}